\documentclass[a4paper,conference]{IEEEtran}
\usepackage[T1]{fontenc}
\usepackage[latin9]{inputenc}
\usepackage{algorithm2e}
\usepackage{amstext}
\usepackage{graphicx}
\PassOptionsToPackage{normalem}{ulem}
\usepackage{ulem}

\makeatletter

\pdfpageheight\paperheight
\pdfpagewidth\paperwidth

\usepackage{cite}
\usepackage{amsmath,amssymb,amsfonts}
\usepackage{algorithmic}
\usepackage{graphicx}
\usepackage{textcomp}
\usepackage{xcolor}
\def\BibTeX{{\rm B\kern-.05em{\sc i\kern-.025em b}\kern-.08em
    T\kern-.1667em\lower.7ex\hbox{E}\kern-.125emX}}

\LinesNumbered
\SetKwRepeat{Do}{do}{while}%

\let\oldnl\nl
\newcommand{\nonl}{\renewcommand{\nl}{\let\nl\oldnl}}

\makeatother

\begin{document}
\title{Organizing Network Management Logic with Circular Economy Principles }
\author{Christos Liaskos, Ageliki Tsioliaridou, Sotiris Ioannidis\\
{\small{}Foundation for Research and Technology - Hellas (FORTH)}\\
{\small{}Emails: \{cliaskos,atsiolia,sotiris\}@ics.forth.gr}}

\maketitle
\textbf{\small{}Abstract\textemdash The traditional cycle of industrial
products has been linear since its inception. Raw resources are acquired,
processed, distributed, used and ultimately disposed of. This linearity
has led to a dangerously low efficiency degree in resource use, and
has brought forth serious concerns for the viability of our natural
ecosystem. Circular economy is introducing a circular workflow for
the lifetime of products. It generalizes the disposal phase, reconnecting
it to manufacturing, distribution and end-use, thus limiting true
deposition to the environment. This process has not been extended
so far to software. Nonetheless, the development of software follows
the same phases, and also entails the use\textendash and waste\textendash of
considerable resources. This include human effort, as well as human
and infrastructure sustenance products such as food, traveling and
energy. This paper introduces circular economy principles to the software
development, and particularly to network management logic and security.
It employs a recently proposed concept\textendash the Socket Store\textendash which
is an online store distributing end-user network logic in modular
form. The Store modules act as mediators between the end-user network
logic and the network resources. It is shown that the Socket Store
can implement all circular economy principles to the software life-cycle,
with considerable gains in resource waste.}{\small\par}

\section{Introduction}

\label{SECINTRO}

The development of computer software is a business of speed. As the
real world gets mapped to the virtual one, more and more concepts,
applications, tools, libraries and techniques appear and disappear
in a matter of months or few years at best. Naturally, any software
developer can see a pattern: the wheel gets re-invented and re-implemented
periodically, spawning torrents of short-lived apps in the process.
If we study the sub-field of network management and control software,
the situation is further aggravated by the fast-paced changes in the
underlying hardware platforms. The Cloud, 5G, virtualization, resource
slicing and IoT are but a few of the terms that where introduced in
the past few years.

Expectedly, surveys identify that a $66\%$ percentage of queried
IT specialists acknowledged difficulty in keeping up with advances
in the networking field~\cite{Loeb.2016}. We can infer that this
ratio can be higher for application developers that do not specialize
in networks. Under these conditions and the push for fast production
cycles, we can safely assume that the ecological planning of resources
is wither downplayed or completely disregarded.

The consequences of roughly designed software has direct and indirect
ramifications to the economy of resources. Directly, the produced
software manages the modern world. If the produced network code is
energy-consuming, error-prone and insecure it will not only waste
network resources, but will also cause resource waste in the real
world. Indirectly, the software development process is not for free.
Developers need to devote person-months for a single product aspect.
Human sustenance resources must be expended in the process, including
food, transportation and climate conditioning. Therefore, bad or repeated
software design translates to waste of such resources at the very
least. Moreover, the supporting network and computing infrastructure
is always on, leading to energy and maintenance resource consumption.

This work proposes the use of the Socket Store as a means of implementing
circular economy in the computer networking software market~\cite{LiaskosSTORE.2017}.The
Socket Store is an online repository of end-client network logic modules.
Within the Store, network providers expose the infrastructure capabilities,
and researchers publish reusable, end-client software modules that
operate on top of them. Developers purchase access to modules fitting
their application, and simply invoke them transparently via a simple,
Berkeley sockets-inspired interface~\cite{Stevens.2004}.

The Store can transform the software life-cycle by altering the disposition
phase as follows:
\begin{itemize}
\item \uline{From disposal to re-acquirement}. Modules in the Store are
used, evaluated, ranked, and commented upon. This enables subsequent
buyers to make more educated choices in network modules incorporated
to their software application.
\item \uline{From disposal to re-processing}. Modules in the Store can
be freely and intuitively re-combined to create new ones that meet
specialized application needs.
\item \uline{From disposal to re-distribution}. The Store remains a distribution
point for all marketed modules. Disposal pertains only to modules
that have proven inefficient and their contributor failed to maintain
them properly.
\end{itemize}
These three general alternatives to the disposal of products are the
pillars of circular economy, a new concept which makes a case for
ecological product design~\cite{tukker2015product}. According to
it, the design refers to the complete life-cycle of the product and
its effects on the environment, rather than just to its intended end-functionality.
As described in the following, the Socket Store allows for the aforementioned
alternative approaches to disposal, constituting an ideal vessel for
enforcing circular economy to software development.

\begin{figure*}[!t]
\begin{centering}
\includegraphics[width=0.88\textwidth,height=4.8cm]{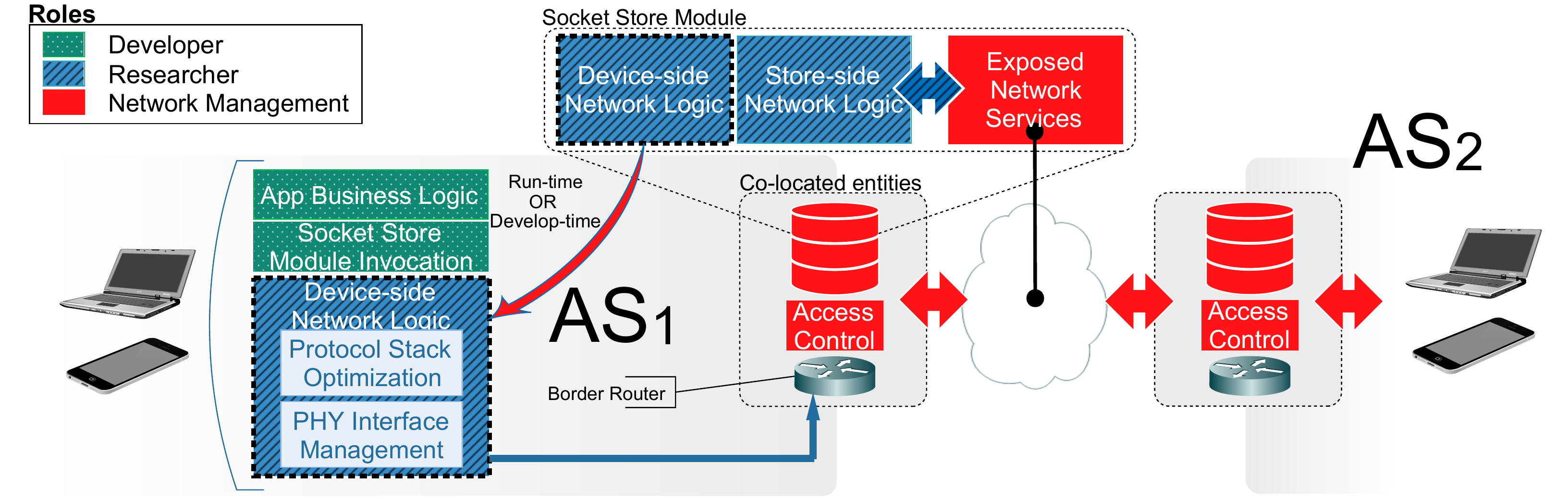}
\par\end{centering}
\caption{\label{fig:OverviewSS}Overview of the Socket Store operation principles
and associated roles.}
\end{figure*}

\section{The Socket Store Concept }

\label{sec:The-Socket-Store}

\textbf{Overview.} Figure~\ref{fig:OverviewSS} presents the components
comprising the Socket Store concept, their general placement within
the network and the jurisdiction they fall into. The description of
the Socket Store will revolve around three involved roles, namely
the mobile/desktop app \emph{developer}, the networking specialist
(\emph{researcher}) and infrastructure provider (\emph{network management},
e.g., ISPs).

The Socket Store proposes a clear separation of roles pertaining to
the programming of end-devices (mobile phones, laptops, desktops,
sensors, etc.). The developer remains responsible for designing and
implementing the business logic (i.e., intended function) of an application.
Communication capabilities required by the application business logic
are delegated to Socket Store modules. Access to these modules is
obtained (\emph{purchased}) by the developer during the implementation
of an application, in a manner reminiscent of app stores (e.g., Google
Play). In other words, we assume the existence of a module-browsing
front-end (e.g., a web-based GUI), where the developer can search
for and obtain modules. Purchased modules are invoked by the developer
via an interface based on the well-known Berkeley sockets. A module
offers sophisticated functionality, security or performance, without
further coding burden for the developer.

The modules can be considered as the agents that are responsible for
mediating between an app and the network resources. We use the term
resources to describe any programmable or parametric network function,
exemplary ranging from packet schedulers, congestion protocols and
flow path deployment to advanced QoS, middlebox access, network state
monitoring and on-demand creation of virtual infrastructure. The Socket
Store is essentially the distributed software platform that hosts
modules and handles their lifecycle, comprising construction, instantiation,
distribution and destruction.

The modules are designed and implemented by researchers. Their scope
is to provide a communication functionality while optimizing a clearly
specified performance objective. The offered functionality can range
from generic (e.g., packet jitter control) to specialized behavior,
e.g., real-time high-quality multimedia transmission. The performance
objective can exemplary express QoS requirements (e.g., packet latency,
loss rate, multimedia quality), device resource expenditure (e.g.,
CPU or battery quotas) and allotted network resources. Moreover, the
network-side optimization can encompass network-wide concerns (e.g.,
load-balanced access to network resources), apart from user-specific
optimizations. The publication of new modules to the Store is intended
to follow the scientific publication process: eponymous submission
followed by a revision round by experts and an open commenting/ranking
system.

\textbf{Module design.} A Socket Store module comprises two separate
but closely communicating components, responsible for \emph{device-side}
and \emph{network-side} optimization respectively. The device-side
component is responsible for carrying out advanced processing of packets
and data flows (e.g., scheduling, adaptive rate control, backup and
fail-over), perform protocol stack optimization (e.g., choose between
IPv4 and IPv6 in terms of latency~\cite{D.Schinazi.2015}) and manage
multiple physical interfaces (e.g., cellular and WiFi) for load-balancing,
redundancy or QoS~\cite{Florissi.2001}. The device-side component
is downloaded to client devices during development, or automatically
upon app installation. In the first case the developer is simply supplied
with a code library during the app development. The second case can
optionally employ a Transferable Object (TO) paradigm (detailed in
the next subsection)~\cite{ObjectManagementGroup.2016}, which mimics
the way app updates are distributed from App Stores. This approach
is intended to allow for transparent network module updates, localization
(e.g., per Autonomous System attributes) and minimal developer coding
burden.

Upon invocation (e.g., on app start-up), the device-side component
contacts the Socket Store in order to create an instance of the corresponding
network-side component(s). The network-side component then allocates
the necessary network resources in communication with the network
management. In a Software-Defined Networking environment for instance,
the network-side component of the Socket Store module can proceed
to setup app-specific routing by interacting with the corresponding
controller. Another example is the creation of Network Virtual Function
chains that are required by the app business logic. Upon app termination,
the network-side component releases the allocated resources (e.g.,
back to a resource pool), while the device-side component frees client-side
resources.

Notice that a module may not implement both the device-side and the
network-side component, but rather adopt a default instead. This capability
allows for modules that target exclusively device-side or network-side
optimization. The default device-side component simply redirects the
app traffic to a Store. The invoked network-side component then acts
as a mediator for the traffic, meeting a network-side optimization
objective. In absence of a network-side component (i.e., device-only
optimization), the app traffic simply defaults to not receiving special
treatment from the network.

\textbf{Hosting and access.} The network management is responsible
for: i) physically hosting the Socket Store, ii) performing access
control to the Store, and iii) exposing interfaces that allow for
the interaction between the modules and the network. Given that the
Socket Store may act as connectivity mediation point for end-user
apps, its physical position within the network is significant. In
the inter-AS case, the Stores should then ideally be co-located with
AS border routers. Since the app traffic needs to cross the AS borders
in any case, co-locating Stores and border routers yield no additional
path hops. In the intra-AS case, Stores should be placed at central
topology points for similar reasons.

Finally, we outline the scope of the Store access control. Given that
modules are purchased by developers, there may be a need to license
and authorize their use from the end-users. In the general case, the
modules are distributed in the context of an app, either directly
or as in-app purchases. Therefore, module access authorization can
be delegated to mobile App Stores. Thus, owning an app also grants
access to the corresponding Socket module(s). It is noted that licensing
per device is also possible in principle. In this case, the device-side
module components tie their operation to physical device characteristics
by enforcing Physical Unclonable Functions (PUFs)~\cite{PUFs.2007}.
The Socket Store access control is also the point for incorporating
defense schemes against Denial-of-Service attacks targeting the Socket
Store.
\begin{figure}
\begin{centering}
\includegraphics[width=0.94\columnwidth]{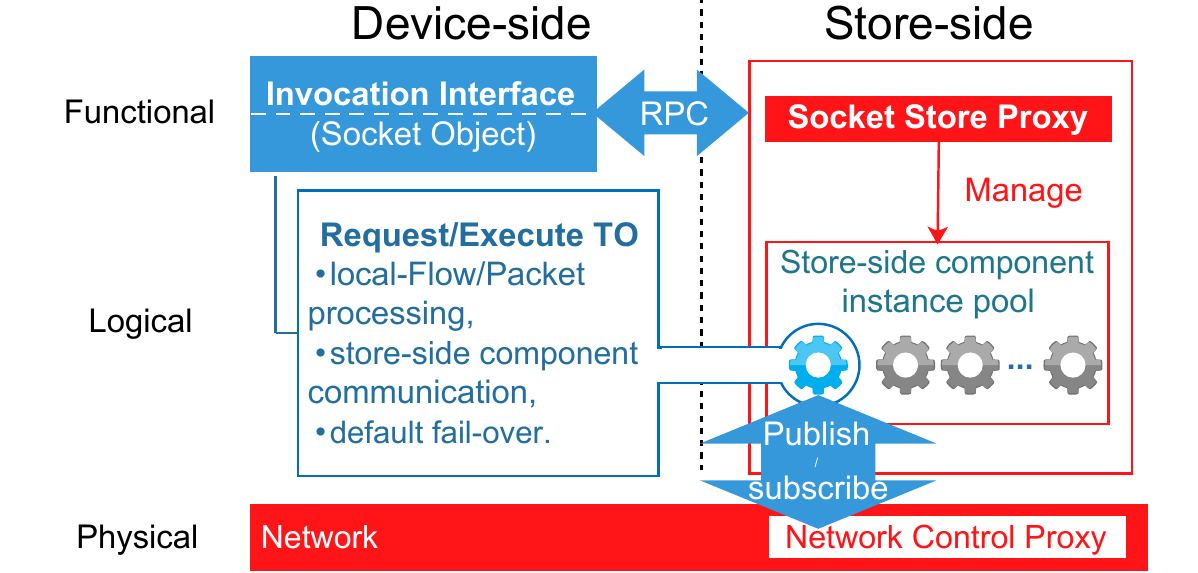}
\par\end{centering}
\caption{\label{fig:arch}Overview of the Socket Store components and communication.}
\vspace{-10bp}
\end{figure}

\subsection{Architectural Aspects\label{subsec:Implementation-Aspects}}

We proceed to study the structure of the Socket Store module components,
the exposed network services and their interconnection.

The system modules follow an object-oriented approach, while their
inter-communication employs the well-known Remote Procedure Call (RPC)
and Transferable Object (TO) patterns~\cite{OMG.2015}. As shown
in Fig.~\ref{fig:arch}, the device-side component is an instance
of the Device-side Socket Object class (DSO), which exposes a functional
interface to the app developer. The DSO can be installed to the device
in two ways. First, it can be received as an offline library during
the app development and be distributed with the app. Second, it can
be downloaded during the first-time initialization of the app (run-time).
In this case, the DSO requests and receives the device-side network
logic from the socket store in the form of a Transferable Object.
A TO (also known as valuetype) is not a remotely referenced object,
but is rather obtained by value and resides at the device-side~\cite{CORBAvaluetype}.
The TO is subsequently stored locally for future use (until an update
occurs). Its execution instantiates the device-side network logic
of the Socket Store module, which encompasses device-local packet/flow
processing and contacting the network for the allocation of resources.
The TO logic is not intended to follow a strictly defined workflow.
For instance, a DSO may run on top of multi-path TCP capabilities
offered by the underlying device operating system, or implement this
functionality itself via proper management of a plain, Berkeley socket
pool~\cite{Florissi.2001}. The DSO destructor signals the network
to de-allocate resources. In case of failure during the TO acquisition
or execution, the DSO can default to the normal network behavior (e.g.,
instantiate a plain Berkeley socket to the intended destination).

The Store-side component is also an instance of a socket object class
(SSO). The SSO instances are created by the Store, after an invocation
from the device-side code. Their role is to interact with the exposed
network services, allocate resources (such as paths) and react to
their modification. The SSO instances are managed by the Store Proxy,
following existing distributed object management approaches~\cite{ZooKeeper}.
Should a DSO signal its destruction (or has remained inactive for
a given timeout), the Proxy returns the SSO instance and its associated
network resources to a pool for re-use, or destroys the SSO instance
and frees the related network resources. The decision is meant to
be adaptive, depending on the popularity of the Socket module in question.

The communication between the DSO and SSO, as well as between the
SSO and the network control follows an event-driven approach, implemented
on top of Remote Procedure Call pattern. The DSO and SSO mutually
expose interfaces to their internal functions. The form of these interfaces
can be defined by the researcher, who is responsible for designing
both components. Moreover, the related functions should be designated
as private and final, to denote that their callback or modification
from the developer is not intended. The communication between the
SSOs and the Network Control is also event-driven, following a publish-subscribe
approach. Particularly, the Network Control publishes events pertaining
to the status modification of network resources, e.g., a change in
the end-to-end latency of an allocated path. An SSO subscribes to
the event types related to its operation and implements the handling
code that should be executed once they are triggered. At the network
layer, the RPC-related signals can receive special treatment from
the network. Next, we detail the structure of DSOs and SSOs.

\textbf{The DSO structure}. In software programming terms, the DSO
is a class instance whose basic structure and functionality is given
in Table~\ref{tabClient}.
\begin{table}[t]
{\small{}\includegraphics[width=1\columnwidth]{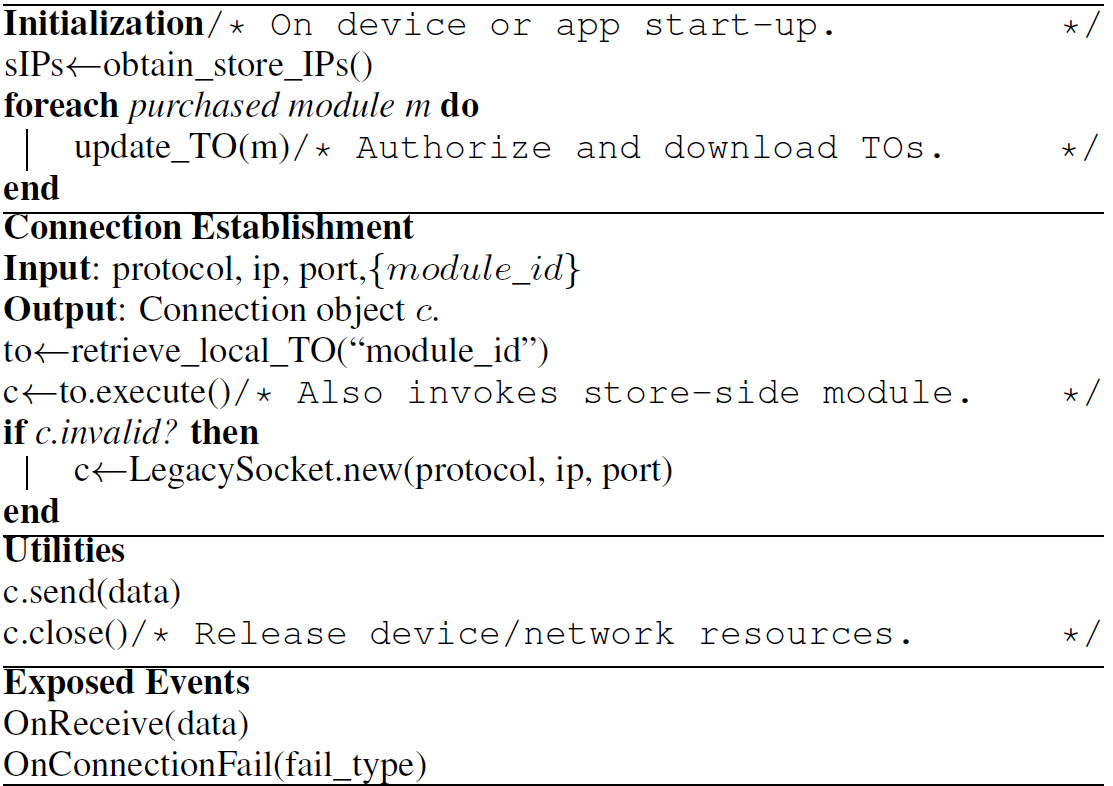}}{\small\par}

\caption{\label{tabClient}{\footnotesize{}Socket module basic invocation interface
and functionality at the device-side (DSO).}}
\end{table}
During the \emph{initialization} phase (app run-time), a DSO retrieves
the Socket Store network identities (IPs). Subsequently, the DSO proceeds
to check for TO updates, in which case it obtains (after proper authorization)
and locally stores them. In order to minimize the overhead of this
process, the TO updates can occur at the same time as app updates,
e.g., periodically or per OS start-up. \emph{Connection establishment}
comprises the retrieval of the corresponding TOs from the local storage
and their execution, which returns a socket connection object. We
note that a string identifier (``module\_id'') is supplied alongside
the usual connection parameters (protocol, port, ip) as an optional
argument. This identifier, which is produced by the Store, corresponds
to a specific parameterization of the related SSOs. Its goal is to
hide unneeded parameterization complexities from the developer. In
case of any error, the connection effort reverts to establishing a
plain Berkeley socket (LegacySocket) instead. The outcome of the connection
can be handled within the DSO code via type checking. (I.e., checking
if the connection attempt returned an object of the class LegacySocket).
Finally, notice that the outlined logic also holds for binding a local
port/interface for receiving incoming connections (apart from connecting
to a remote destination).

The returned connection object offers a Berkeley socket-inspired interface
at a minimum (e.g., supporting the $\texttt{send()}$, $\texttt{close()}$
functions etc.), which constitutes common knowledge to developers.
Moreover, a DSO can offer additional \emph{utilities,} depending on
the objective of the module as a whole. For example, a multimedia
streaming-oriented module may overload the original send function
to directly stream video files, e.g., $\texttt{send(AVIFileStream)}$.
Another example could be a Delay Tolerance-oriented module~\cite{DTN.2011},
which adds functionality for $\texttt{UndoSend()}$, $\texttt{RedoSend()}$.
Lastly, the DSO exposes events, such as incoming data, failure to
uphold a performance objective, etc., which should be handled by the
developer.
\begin{figure}
\begin{centering}
\includegraphics[width=0.82\columnwidth]{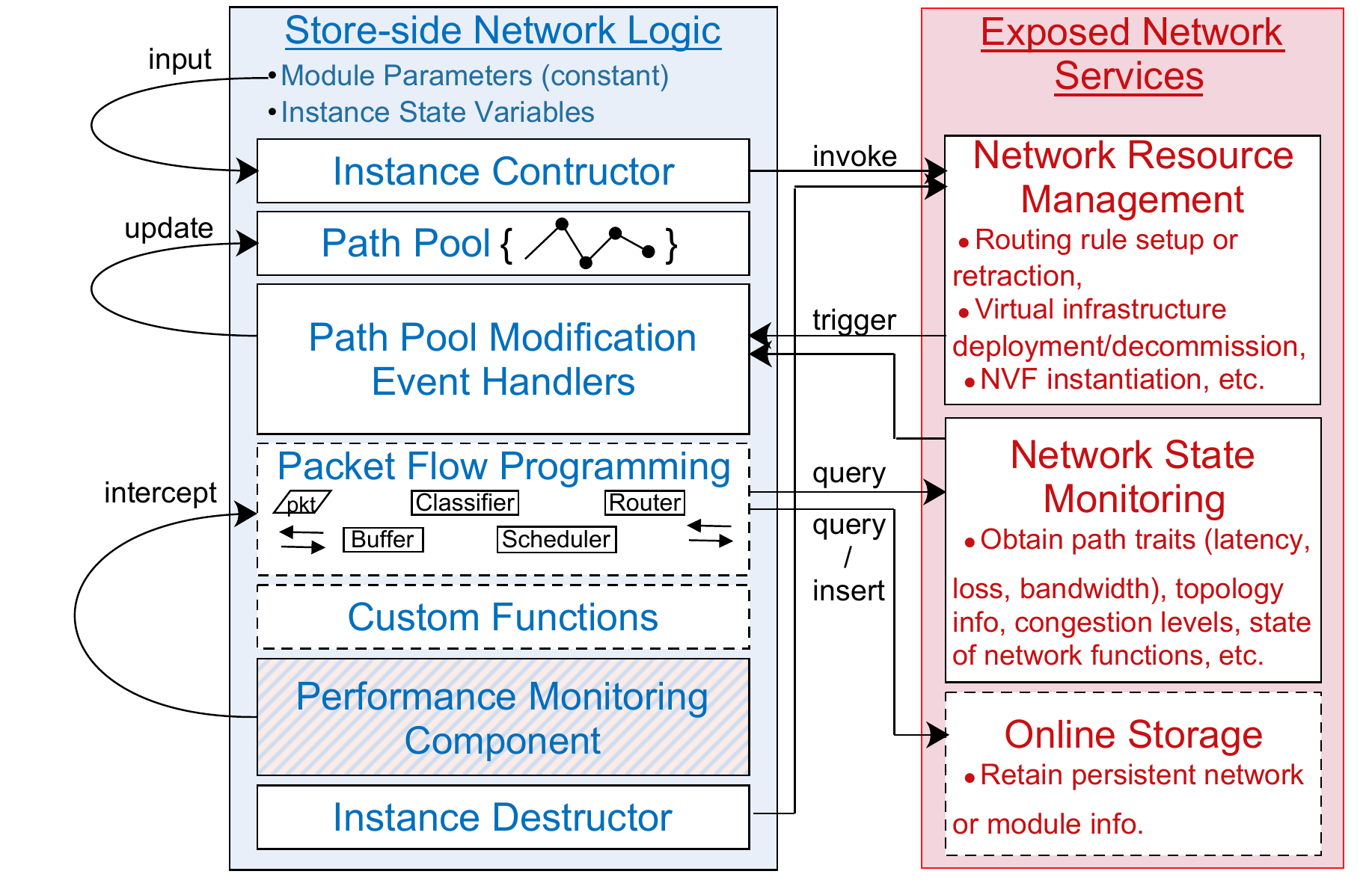}
\par\end{centering}
\caption{\label{fig:store-side}Schematic of the components comprising the
store-side network logic of a module instance (SSO), and their relation
to network services.}
\end{figure}
\textbf{The SSO structure}. The schematic of the SSO structure and
functions is given in Fig.~\ref{fig:store-side}. SSO also follows
an object oriented approach, with a constructor, a destructor, member
functions, parameters and variables, which can be marked as private
or public, etc. The parameters are represented as \emph{constants},
and each separate parameterization is mapped to the string identifier
``module\_id'' described earlier. The instance variables are accessible
from the SSO member functions and are mainly responsible for expressing
the internal state of the SSO instance. The SSO constructor is responsible
for contacting the network control for allocating required resources
(e.g., path, VPN, firewall deployment). The SSO destructor is responsible
for de-allocating these resources when triggered by the DSO (or after
a timeout).

During its development, the SSO subscribes to the network events that
it should react to. The Network Control (state monitoring service)
is responsible for publishing these events, eventually triggering
the subscribed SSO event handlers. An example is given in Fig.~\ref{fig:store-side}
in the context of the $\text{\texttt{PathPool}}$ programmatic facility.
$\text{\texttt{PathPool}}$ is a high-level collection of $\text{\texttt{Path}}$
objects, each corresponding to a path allocated to the socket module.
The $\text{\texttt{Path}}$ object facilitates programmatic query
of path traits such as latency and bandwidth of intermediate links.
Should, e.g., the latency of a link change, the network monitoring
service triggers the corresponding subscribed events at the Store.
The SSO path event handler is triggered if its $\text{\texttt{PathPool}}$
is affected by the change in the link state.

SSO can operate at flow-level granularity, in which case it operates
only when the app-specific network routing should change. Nonetheless,
advanced socket modules (e.g., novel schedulers) may require processing
at pack-level granularity. The SSO supports this functionality by
supporting packet-flow programming~\cite{FBP.2010}. Nonetheless,
packet-level processing is generally computationally intensive, e.g.,
requiring the processing of packets at user memory space~\cite{FreeBSDManPages.2004},
or via programmable hardware (e.g., FPGA~\cite{salim2014programmable}).
Therefore, such functionality should be considered when pricing a
socket module. In both cases, however, the network management should
provide facilities for module \emph{performance monitoring} (e.g.,
libraries for monitoring the end-to-end latency). The module monitoring
is required for deducing whether its performance objective is being
met, and also for objectively ranking competing socket modules.

Finally, we outline inter-Store and Store-Cloud interactions. Inter-store
communication can employ modules, i.e., each SSO can also act as a
module client and call upon other socket modules. SSOs can also interact
with online storage services to facilitate the module operation. For
instance, a DTN-oriented socket may store pending packets in the Cloud,
until the connection end-point becomes available.

\section{Potential and Challenges}

\label{sec:Discussion}

The Socket Store allows for a clearer separation of concerns among
app developers, researchers and network providers. Apart from the
envisioned motives mentioned in Section~\ref{SECINTRO}, the Store
concept can provide additional potential. Firstly, it could bring
the functionality of novel networking paradigms to the end-users.
For instance, socket modules could provide a Named-Data Networking
interface to developers~\cite{NDN.2014}, allowing for an ``Interest''
and ``Data''-based communication while hiding underlying network
operations. A similar approach can be followed for the Recursive Inter-Network
Architecture (RINA) \cite{Wang.2014}, which models networking as
inter-process communication. The Store could also serve as an end-user
access point to novel schemes that aim at providing QoS in inter-AS
routing~\cite{VasileiosKotronis.2014}. Secondly, the Store provides
the network management with direct knowledge of the users' requirements
and intentions, limiting the need for indirect monitoring and enabling
network-wide resource optimization. Lastly, the proposed scheme can
constitute a platform for visibility and fair comparison of research
contributions. Automatic evaluation and ranking of novel schemes is
possible, exploiting existing public testbeds in the process~\cite{ofelia.2014}.
The main questions in our future research agenda are:

\textbf{Interface optimization and evolution.} Apart from the described
approach, is there a more efficient way for interfacing among devices,
the Stores and the Network? What would be the decisive performance
metrics? Can we ensure inter-Store compatibility and interface extensibility?

\textbf{Developer usability.} How can a developer search for a module
that fits his app needs easily and effectively? Is there an effective
way of bringing this facility directly inside app development environments?

\textbf{Market model and pricing.} What should be the purchase and
pricing model for socket modules? Purchase access once per app (or
enable as in-app purchase), or charge per usage? How can we monetize
the value of network resource access in a rational and automatic manner?

\textbf{Deployment.} Socket Store instances are required in the vicinity
of clients. What network-intermediate points of deployment are needed
(e.g., at IXPs)? What factors should drive an incremental deployment
strategy?

\textbf{Security.} Which are the new attack vectors that are introduced
by the Socket Store? How can we enforce security and privacy in the
communication between the involved entities?

\section{Related work}

\label{sec:Related-work}

Building network-aware apps and application-aware networks has constituted
an early, notable research goal~\cite{Zurawski.2013}. Nonetheless,
to the best of the authors' knowledge, there has not been an effort
to make research knowledge modular and reusable, with a consistent
interface between the network and the applications.

\textbf{Network-side.} In 1998, Lowekamp et al. proposed the REMOS
unified network query interface, which allowed applications to request
network topology and congestion information, in order to tune their
behavior accordingly~~\cite{Lowekamp.1998}. PerfSONAR is a more
recent approach towards this direction~\cite{Zurawski.2013}, employing
a more standardized metadata format for describing network characteristics~\cite{NMWG.2006}.
The Unified Network Information Service~\cite{ElHassany.2013}, offered
an alternative solution with scalability, security and performance
benefits. Application domain-specific network querying solutions exist
as well, such as MonALISA for distributed systems~\cite{Legrand.2009},
and the Network Weather Service for meta-computing~\cite{Wolski.1999}.
These approaches also employ standardized formats for representing
the network status and configuration~\cite{YANG.2010}. In Socket
Store terminology, these works are means of exposing the network infrastructure
capabilities, allowing researchers to build Socket modules on top
of them.

Novel infrastructure capabilities, such as Infrastructure-as-a-Service,
can be exposed in reusable component form at the Store via the NFV
paradigm~\cite{ETSI.2013}. Additionally, the Store can benefit from
directly interfacing with SDN technology, which already offers the
necessary interfaces to query and modify the network state in a modular
manner (e.g., network \emph{intents})~\cite{GemberJacobson.2014}.
Based on the advances in SDN and NFV, recent works have proposed new
paradigms that can enable QoS guarantees across the Internet, crossing
the Autonomous System borders~\cite{Gupta.2014}.

It is also noted that the ``store'' term is also used for exposing
network functions and services in a modular form (e.g., NFV stores)~\cite{MasergyCorp.2016}.
There also exist user-friendly online platforms for chaining such
modules~\cite{CanonicalLtd.2016}. Since these works expose the network
capabilities in an easy-to-use form, they also promote their interfacing
with the client-side logic at the Socket Store. It is noted that network
functions should preferably be exposed by the network providers and
not by independent parties, to avoid code trust issues~\cite{Maimonis.2015}.

\textbf{Device-side.} Given that the Berkeley Socket API is popular
among developers~\cite{Stevens.2004}, researchers sought to preserve
its principles while hiding complex network logic underneath it. Protocol
optimization, packet scheduling and hardware interface selection have
constituted notable optimization goals~\cite{Florissi.2001}, especially
for QoS monitoring and provision in mobile devices. Apple is also
incorporating a transparent socket selection between IPv4 / IPv6 in
its most recent OS revision~\cite{D.Schinazi.2015}. Apart from device-side
optimizations, there have been proposals for sockets that signal the
network for the type of treatment they require~\cite{Welzl.2011}.
Similar approaches have been proposed for the transparent evolution
of the TCP protocol~\cite{Nabi.}, and multipath TCP and HTTP constitute
notable cases~\cite{Kim.2014}. The recently started IETF Transport
Services working group targets the API standardization of socket augmentations~\cite{Fairhurst.2015}.

Within the Socket Store, such works can be freely published as reusable
modules and become accessible to developers. This access can be readily
uniform and transparent, with trivial coding overhead~\cite{ObjectManagementGroup.2016}.
The Berkeley API is retained as the basic interface. Finally, the
Socket Store is the point where the two ends-the device and network
states-can meet, allowing end-users to benefit from their interaction.

\section{Conclusion}

\label{sec:Conclusion}

The present work proposed the use of circular economy principles in
software design. To this end, it proposed the use of the Socket Store,
a novel approach for realizing network-aware applications and application-aware
networks. The Socket Store makes network logic available for purchase
by developers, who can combine and incorporate them to their applications.
It naturally acts as a platform that enforces software re-distribution,
re-processing and re-use. These traits align the Store perfectly to
the principles of circular economy, making for educated, ecological
and proper use of human and raw resources, while allowing for fast-paced
software development.

\section*{Acknowledgment}

This project was funded by the European Union via projects ``CE-IoT:
A Framework for Pairing Circular Economy and IoT'' (Marie Sk\l odowska-Curie
RISE action, GA EU777855) and ``VISORSURF: A Hardware Platform for
Software-driven Functional Metasurfaces'' (Future Emerging Topics
- FETOPEN-RIA, GA EU736876).


\end{document}